\documentclass[10pt,letterpaper]{article} 
\usepackage[english]{babel}
\newcommand{\s}{\\[1ex]}
\newcommand{\goesto}{\scriptstyle\rightarrow}
\newcommand{\beq}{\begin{equation}} 
\newcommand{\eeq}{\end{equation}}
\newcommand{\lbl}{\label}
\newcommand{\q}{\quad}

\newcommand{\arrow}{\rightarrow}

\newcommand{\non}{\nonumber}

\newcommand{\re}[1]{(\ref{#1})}

\newcounter{examplenum}[section]
\newcounter{dignum}[section]
\newcounter{exercisenum}
\newcommand{\bx}{\vrule height8pt width3pt depth 0pt} 
\newenvironment{proof}{\bf Proof: \rm}{$\bx$ \s}

\newtheorem{theorem}{Theorem}
\newtheorem{lemma}[theorem]{Lemma}

\newtheorem{proposition}[theorem]{Proposition}

\begin{document} 
\title{Asymptotics of a proposed delay-differential equation of motion for charged particles} 
\author{Stephen Parrott \\ 
2575 Bowers Rd.\\
Gardnerville, NV 89410\\
Email:  spardr@netscape.net
}
\maketitle
\noindent
Running head:  Asymptotics of a delay-differential equation of motion
\s
\noindent
%Abstract:
%\begin{quote}
%\small
\begin{abstract}
F.\ Rohrlich has recently published two papers advocating 
a particular delay-differential (DD) equation as 
an approximate equation of motion for classical charged particles,  
which he characterizes as providing a 
``fully acceptable classical electrodynamics''. 
We study the behavior in the remote past and future of
solutions of this equation 
for the special case in which the motion is
in one spatial dimension. 

We show that if an external force is applied for a finite time,
some solutions exhibit the property of ``preacceleration'',
meaning that the particle accelerates {\em before} the force is applied,
but that there do exist solutions without preacceleration.
However, most solutions without preacceleration 
exhibit ``postacceleration'' into the infinite future, meaning that 
the particle accelerates {\em after} the force is removed.  
Some may regard such behavior as sufficiently ``unphysical''
to rule out the equation.

More encouragingly, analogs of the pathological ``runaway'' solutions 
of the Lorentz-Dirac
equation do not occur for solutions of the DD equation. 
We show that when the external force eventually vanishes,
the proper acceleration vanishes asymptotically in the future, 
and the coordinate velocity becomes asymptotically constant.
\end{abstract}
%\end{quote} 
%\pagebreak[4]
%
\noindent
\section{Introduction}
The correct equation to describe the motion of a charged particle
in flat spacetime (Minkowski space)
has long been a matter of speculation and controversy.
The most-mentioned candidate has been the Lorentz-Dirac equation,
written here for units in which light has unit velocity
and metric tensor of signature $[+,-,-,-]$ :
\beq
\lbl{ldeq}
m \frac{du^i}{d\tau} = q {F^i}_\alpha u^\alpha + \frac{2}{3} q^2
\left[\frac{d^2 u^i}{d\tau^2} + \frac{du^\alpha}{d\tau}
\frac{du_\alpha}{d\tau} u^i
\right]
\eeq
This is written in traditional tensor notation with repeated indices 
summed and emphasized in Greek;
$u = u^i$ denotes the particle's four-velocity, $m$ and $q$ its mass
and charge, respectively, $\tau$ its proper time,
and $F_{ij}$ the (antisymmetric) tensor describing an external electromagnetic
field driving the motion. 

However, many objections have been raised to this equation.
%References to typical literature and discussion too extensive to be given here 
%can be found in \cite{parrottbook} and \cite{parrott2}. 
Among them are the existence of ``runaway'' solutions
in which the acceleration increases exponentially with proper time
even when the external field asymptotically vanishes.
In some physically reasonable situations, {\em all} solutions are runaway  
(Eliezer, 1943; for later references see Parrott, 1987, Section 5.5).
%\cite{eliezer}, \cite{parrottbook}, \cite{parrott2}.  
Even in favorable cases in which non-runaway solutions exist,
they may exhibit so-called ``preacceleration'' in which the particle
begins to accelerate {\em before} the external field is applied
(Rohrlich, 1965).
%\cite{rohrlichbook}.

F.\ Rohrlich 
(1997, 1999)
%\cite{rohrlichamj} \cite{rohrlichpr} 
has recently advocated a new equation of motion which
(Rohrlich, 1999) claims without proof 
``[has] no pathological solutions''.  
This new equation is a delay-differential equation given below,
which we shall call the ``DD equation''.
The present paper derives from a study of the DD equation with the aim
of determining the correctness of such claims.%  

(Rohrlich, 1997) claims  
that the DD equation has neither preaccelerative nor ``runaway'' 
solutions.
We shall show that for the case of a nonzero external force 
applied for only a finite time,
the DD equation does
admit preaccelerative solutions, and that 
these are all ``runaway'' in the weak sense that 
the acceleration does not vanish asymptotically in the distant past
as one would expect.
However, this does not rule out the DD equation as a realistic equation of motion
because we also show that these preaccelerative solutions can be eliminated 
by appropriate choice of generalized initial conditions (defined below).
But then another problem arises:
assuming this choice, most solutions exhibit ``postacceleration'',
meaning that the acceleration persists after the external field is 
turned off.  

Postacceleration is not as bad as preacceleration because there is no
violation of causality, but 
postaccelerative solutions could be considered pathological.
Suppose we are sitting in a room shielded from electromagnetic fields
watching a beam of identical charged particles shoot in the window.
It might seem strange if some of the particles 
speeded up, while others slowed down, for no apparent reason,
according to their past histories.  
This is what the DD equation predicts.

	(Rohrlich, 1997) presents the DD equation 
as a modification of the following equation
proposed by (Caldirola, 1956): 
%\cite{caldirola}:  
\beq
\lbl{caldirolaeq1}
\frac{m}{\tau_0} 
[ u^i(\tau-\tau_0) - u^\alpha (\tau-\tau_0) u_\alpha(\tau) u^i(\tau) ] 
+ \frac{q}{c} {F^i}_\alpha u^\alpha (\tau) = 0
\q.
\eeq 
This is written in units in which the velocity of light is $c$,
and $\tau_0 := 4q^2/(3mc^3) \approx 1.2 \times 10^{-23} \mbox{sec}$
is a constant with the dimensions of time.
With appropriate units employing our convention 
that the velocity of light is unity,
this can be written: 
\beq
\lbl{caldirolaeq2}
\frac{m}{\tau_0} 
[ u^i(\tau-\tau_0) - u^\alpha (\tau - \tau_0) u_\alpha (\tau) u^i(\tau) ] 
+ q {F^i}_\alpha u^\alpha = 0
\q.
\eeq
%The actual transformation laws are more complicated than merely replacing
%$c$ by 1 in \re{caldirolaeq1},
%but nevertheless \re{caldirolaeq2} results if 
%units of charge are unchanged, and units of mass
%are changed consistent with  the change in velocity of light 
%(Parrott, 1987, Appendix on Units).
%(\cite{parrottbook}, Appendix on units). 

The motivation presented by 
(Caldirola, 1956) 
%\cite{caldirola}
involves starting with the Lorentz equation (the classical force equation
ignoring radiation reaction), 
\beq
\lbl{lorentzeq}
m \frac{du^i}{d\tau} = q {F^i}_\alpha u^\alpha 
\q,
\eeq
and attempting to replace $du^i/d\tau$ by the difference quotient
\beq
\lbl{diffquot}
\frac{u^i(\tau) - u^i (\tau-\tau_0)}{\tau_0}
\q.
\eeq
But since both sides of \re{lorentzeq} are orthogonal to $u$,
while \re{diffquot} need not be,
one ought to project \re{diffquot} into the subspace orthogonal
to $u(\tau)$.  This projection will kill any multiple of $u(\tau)$,
so the resulting equation can be rewritten as \re{caldirolaeq2}  

Rohrlich (1997)
%\cite{rohrlichamj} \cite{rohrlichpr} 
suggested the following modification of \re{caldirolaeq2}.
This is what we are calling the DD equation:
\beq
\lbl{rohreq1}
m_1 \frac{du^i}{d\tau} = 
f^i(\tau) +  
m_2 [ u^i(\tau-\tau_1) - u^\alpha (\tau - \tau_1) u_\alpha (\tau) u^i(\tau) ] 
\q.
\eeq 
\newcounter{rohreqonenum} 
\setcounter{rohreqonenum}{\value{equation}} 
Here $f(\tau) = f^i(\tau) $ is a four-force orthogonal to $ u(\tau)$, 
$m_1$ and $m_2$ are presumably nonzero 
parameters associated with his motivation
of the right side as an approximation to the self-force on a 
spherical surface charge, and $\tau_1$  is a positive parameter.
(The sign of the second term in brackets differs from his because
his metric is opposite in sign to ours.)
To avoid degenerate cases, we will assume below that $m_2$ and 
$m_1$ are nonzero; actually, they would be expected to be positive.

(Rohrlich, 1997) attributes the conjecture of \re{rohreq1} 
to (Caldirola, 1956),
%\cite{caldirola},
but \re{rohreq1}  is not mathematically equivalent to \re{caldirolaeq2},
and (Caldirola, 1956)
%\cite{caldirola} 
did not propose it.  
For example, Caldirola's equation for a compactly supported force
admits periodic solutions (to which Caldirola attached great importance
as possibly describing spin-like internal particle motions),
while the DD equation does not (as the analysis of this paper will make clear). 
Note also that the DD equation \re{rohreq1} is a delay-differential equation,
while Caldirola's equation is a difference equation involving no derivatives. 

The abstract of (Rohrlich, 1999), begins as follows:
\begin{quote}
``The  self-force for the classical dynamics of finite size particles
is obtained. It is to replace the one of von Laue type obtained
for point particles. $\ldots$''%
\footnote{The equation ``of von Laue type'' is the Lorentz Dirac equation
\re{ldeq}.}
\end{quote} 
(Rohrlich, 1999) emphasizes the contribution of 
(Yaghjian, 1992, Appendix D., Equation D.19). 
Yaghjian derived a rest-frame version of the DD equation 
as an uncontrolled approximation%
\footnote{For example, the derivation ignores nonlinear terms.
``Uncontrolled'' describes the fact that no attempt is made to  
quantitatively assess
the magnitude of possible errors which such simplifications might introduce.)
} 
to the self-force on a charged particle modeled as a ``rigid'' sphere.% 
\footnote{
Yahgjian mentions that essentially the same equation was 
``stated without proof'' by (Page, 1918),
and gives references to other precursors.  Also, Caldirola refers
to precursors of his proposed equation.  The history of these ideas
is extremely tangled, and we make no attempt to present all aspects
of it here.}

Yaghjian's equation is written for a particle slowly moving
relative to some given Lorentz frame. 
If the given Lorentz frame is the particle's rest frame
at a given proper time, and it is written in four-dimensional notation,
then it becomes the DD equation.  
Thus the DD equation \re{rohreq1} is a relativistic generalization 
of Yaghjian's equation, and its statement in (Rohrlich, 1997) 
appears to be the first explicit statement in the literature.   
Moreover, (Rohrlich, 1997), goes beyond its mere statement to present it as a
``fully acceptable classical electrodynamics''. 
By comparison, Yaghjian obtained it as an approximation without comment
on the approximation's domain of validity.

Because of this and the above quote from the abstract 
of (Rohrlich, 1999) stating that it ``obtained'' the equation, 
an earlier version of the present paper called the DD equation \re{rohreq1}
``Rohrlich's  equation''.
However, when Professor Rohrlich submitted a report as an identified
referee for that version,
he strongly objected to the name ``Rohrlich's equation'',
attributing it instead to Caldirola and Yaghjian.
In deference to his wishes, we sidestep the convoluted question of the 
origin of equation \re{rohreq1} by employing the neutral name ``DD''
equation.

Although Rohrlich's motivation for the DD equation \re{rohreq1} involves thinking 
of the particle as a charged sphere, the equation of motion itself
is the equation for a moving point, the center of the sphere.
Thus the DD equation is mathematically the equation of a point particle,
and we shall consider it as such; the motivation of the charged sphere  
will not enter into our considerations.

To study \re{rohreq1}, it will be convenient to eliminate inessential constants 
by introducing a new time unit equal to $\tau_1$ old time units
and by using available notational freedom to absorb the two 
parameters $m_1$ and $m_2$ into a single parameter $m$,  
obtaining  
\beq
\lbl{rohreq}
 m \frac{du^i}{d\tau} = f^i + 
 [ u^i(\tau-1) - u^\alpha (\tau - 1) u_\alpha (\tau) u^i(\tau) ] 
\q.
\eeq
\newcounter{rohreqnum} 
\setcounter{rohreqnum}{\value{equation}}
This can be done by dividing both sides of \re{rohreq1} by $m_2$,
and renaming $f$ and $m_1$.  

The resulting equation \re{rohreq}, which we shall call 
the ``normalized DD equation'', is equivalent to the original
DD equation \re{rohreq1} in the sense that knowledge of all solutions 
of either
translates immediately into knowledge of all solutions of the other. 
In response to a referee's skepticism about this,
we added Appendix 2, which works out in detail
the precise relations between the two equations. 
These will also be briefly sketched just below for the reader's convenience.

The analysis of the paper {\em could} be written without introducing
the normalized DD equation  \re{rohreq}, and the conclusions 
would be essentially 
the same, apart from trivial changes of language. 
We chose not to do that because retaining the  original, more complicated
notation would only distract attention from 
the more fundamental ideas which we want to emphasize. 

\newcommand{\taubar}{\bar{\tau}}
\newcommand{\ubar}{\bar{u}}
\newcommand{\fbar}{\bar{f}}

To avoid any misunderstanding, we make explicit that 
the unknown function
``$u$'' in \re{rohreq} is technically not the same as the  
``$u$'' in \re{rohreq1}, but the two ``$u$''s are related in 
a mathematically straightforward and physically transparent way
given below in equation \re{ubar}, 
so that conclusions about solutions of \re{rohreq} can be directly
translated into conclusions about solutions of \re{rohreq1},  
and vice versa. 

More explictly, changing to the new time unit corresponds
to multiplying the old metric by $\tau^{-2}_1$. 
This results in a
new proper time 
$\taubar := \tau_1\tau$ and a new four-velocity $\ubar$ satisfying
\beq
\lbl{ubar} 
\bar{u}(\bar{\tau}) = \tau_1 u(\tau)\q.
\eeq 
Equivalently, for any real number $s$,
\beq
\lbl{ubar2}
\bar{u}(s) = \tau_1 u(\tau_1 s)\q.  
\eeq
This simple relation allows us to translate any conclusion about
$\bar{u}$ into a corresponding conclusion about $u$, and vice versa.

The original DD equation \re{rohreq1} is thus equivalent
to an equation like \re{rohreq} but with symbols renamed: 
\beq
\lbl{rohreqbar}
 m \frac{d\ubar^i}{d\taubar} = \fbar^i + 
 [ \ubar^i(\taubar-1) - 
\ubar^\alpha (\taubar - 1) \ubar_\alpha (\taubar) \ubar^i(\taubar) ] 
\q.
\eeq
\newcounter{rohreqbarnum} 
\setcounter{rohreqbarnum}{\value{equation}}
The details of the relations of the renamed symbols to the originals
are given in Appendix 2. 

In the rest of the paper, we omit the bars for typographical
simplicity, and analyze \re{rohreq} instead of \re{rohreqbar}, 
leaving to the reader the straightforward translation of mathematical results
about \re{rohreq} to conclusions about \re{rohreq1}.
All of our conclusions about \re{rohreq1}  
will be essentially identical to 
the corresponding conclusions about \re{rohreqbar}, apart from 
trivial changes in language or notation.  
For example, Theorem 5's conclusion 
that for eventually vanishing force,
$\lim_{\taubar \goesto 0} d\ubar/d\taubar = 0$, translates into
the conclusion that solutions $u$ of \re{rohreq1} satisfy
$\lim_{\tau \goesto 0} du/d\tau = 0$.  
\section{The DD equation for one-dimensional motion}
We shall study the (normalized) DD equation \re{rohreq} for the case of motion
in one space dimension.  
(Henceforth we omit the reminder ``normalized''.)
Analysis of this special case turns out to be sufficient to answer the
questions motivating this work.  

For this case, we may work in a two-dimensional Minkowski space
with typical vector $x = (x^0, x^1)$ and metric tensor
$g(x,x) = (x^0)^2 - (x^1)^2$.  
Since $u^\alpha u_\alpha = g(u,u) = 1$, we may write
\beq
\lbl{rapiditydef}
u = (\cosh \theta, \sinh \theta)
\q,
\eeq
where this defines the ``rapidity'' parameter $\theta$.  
Define
\beq
\lbl{wdef}
w := (\sinh \theta, \cosh \theta)
\q,
\eeq 
so that $w$ is a spacelike unit vector orthogonal to $u$,
and any vector orthogonal to $u$ must be a multiple of $w$.

In the DD equation \re{rohreq}, all of $du/d\tau$,
$f$, and the bracketed term are orthogonal to $u$,
and hence multiples of $w$.
This can be seen explicitly for the left side:
$$
m\frac{du}{d\tau} = m \frac{d\theta}{d\tau} w 
\q.
$$
It is natural to name 
\beq
\lbl{Adef}
A := \frac{d\theta}{d\tau}
\eeq
the ``scalar proper acceleration''.

Write
$$
f = Ew
\q,
$$
where this defines the scalar $E$.  When 
$f^i = {F^i}_\alpha u^\alpha$ is a Lorentz force, $E$ is the electric field
(nominally relative to the Lorentz frame corresponding to the basis $u, w$
for two-dimensional Minkowski space, but actually relative to the Lorentz
frame determined by {\em any} orthogonal basis, as is revealed by 
straightforward calculation).

The bracketed term of \re{rohreq}, 
being a multiple of $w$, is equal to its projection
on $w$.  The projection of an arbitrary vector $v$ on $w$
is $-v^\alpha w_\alpha w$, so the projection of the bracketed term of \re{rohreq} on $w$ 
is $w$ multiplied by the scalar
$$
-(\cosh \theta(\tau-1)\sinh \theta(\tau) - 
\sinh \theta (\tau-1) \cosh \theta (\tau) = 
\sinh (\theta (\tau-1) - \theta (\tau))
\q.
$$
Projecting the entire DD equation \re{rohreq} on $w$ yields
the equivalent scalar equation
\beq
\lbl{rohreq2}
mA = m \frac{d\theta}{d\tau} = E + \sinh (\theta(\tau - 1 ) - \theta(\tau))
\q.
\eeq
We shall call this the variant of the DD equation for 
one-dimensional motion the  ``DD1 equation''.

Equation \re{rohreq2} may be regarded as a delay-differential
equations of the general form 
\beq
\lbl{delayeq}
\frac{d\lambda}{d\tau} = \Phi ( \tau, \lambda (\tau), \lambda (\tau-1))
\q,
\eeq
\newcounter{delayeqnum} 
\setcounter{delayeqnum}{\theequation}
%
%NOTE:  The correct constuction appears to be 
%\setcounter{delayeqnum}{\value{equation}},
%but this appears to work, and I don't want to change it because
%I don't remember where I used it.
%
with $\Phi$ a given function.  
If we imagine $E(\tau)$ as a given function of proper time $\tau$, 
then \re{rohreq2}
is of the form \re{delayeq} with 
$\Phi (\tau, r, s) := E(\tau) + \sinh (r-s)$.) 

In general, the situation is more complicated because 
$\theta$ is defined by \re{rapiditydef}, with 
$u^i := dz^i/d\tau$, 
where $z^i(\tau) $ represents the particle's wordline. 
Then $E$ is usually not explicitly given as a function of proper time,
but instead is given as a function $E(z, u) = E(z(\tau), u(\tau))$ 
of the Minkowski coordinates and four-velocity, 
with its proper time dependence acquired indirectly from the time-dependence
of the latter.  However, we may still regard the equation 
\re{rohreq2} as of the form \re{delayeq} by imagining solving
\re{rohreq2} for $\theta (\tau)$, which determines $u(\tau)$,  
then (by integration) $z(\tau)$, and finally $E(z(\tau), u(\tau))$.  
(The existence of the solution $u(\tau)$ will be established in the
next section.) 
Following common abuse of notation, we write $E(\tau)$
in place of $E(z(\tau), u(\tau))$.

This shows that $\theta$ does satisfy {\em some} delay-differential
equation of the form \re{delayeq}.
We shall show that this severely restricts the form of the function 
$\theta$.
For example, we'll show that if $\tau \mapsto E(\tau)$ has compact support,
then $\theta$ must be bounded on any semi-infinite interval 
$[\tau_0, \infty)$.  

Similar remarks apply to the equation obtained by differentiating
\re{rohreq2}: 
\beq
\lbl{acceq}
m \frac{dA}{d\tau} = \frac{dE}{d\tau} + 
\cosh (\theta (\tau-1) - \theta(\tau)) [A(\tau-1) - A(\tau)]
\q.
\eeq
This may also be regarded as of the form \re{delayeq}
if we imagine that we have already solved for $\theta$.
At first sight this may seem strange because if we have solved
for $\theta$, then we also have $A := d\theta/d\tau$,
so there is no need to solve \re{acceq} for $A$.
Nevertheless, the observation that $A$ satisfies
an equation of the form \re{acceq} is useful 
because it severely constrains $A$.  For example,
we'll show that it implies that for a force $E(\cdot)$ with
compact support, $\lim_{\tau \arrow \infty} A(\tau)$ exists. 
Then combining this with the above-mentioned fact 
that $\theta$ is 
bounded will imply that in fact $\lim_{\tau \arrow \infty} A(\tau) = 0 $.
\section{General delay-differential equations of form \re{delayeq}}
This section reviews some simple and well-known facts about 
general delay-differential equation of the form \re{delayeq}.  
When discussing such equations, we will always call them 
``delay-differential'' equations, which will never be abbreviated.
We reserve the term ``DD equation'' for the particular special case
\re{rohreq} or one of its variants such as \re{rohreq1}.

Let the function $\Phi$ in the delay-differential equation 
\re{delayeq} be $C^1$ 
(i.e., continuously differentiable). 
Suppose that $\tau \mapsto \lambda(\tau)$ 
satisfies this equation. 
If we regard $\lambda (\tau)$ as given 
on some interval $[n-1, n]$, then \re{delayeq} becomes an ordinary
differential equation for $\lambda$ on $[n, n+1]$ of the form
\beq
\lbl{ordeq} 
\frac{d\lambda}{d\tau} = \Psi (\tau, \lambda (\tau)) 
,   
\q
n \leq \tau \leq n+1
\q,
\eeq
which is covered by the standard existence and unigueness theorems.

This observation reveals the general structure of solutions of \re{delayeq}.
Choose $\lambda(\cdot)$ to be an arbitrarily chosen $C^1$ function
on any closed interval of length 1,
say the interval $[0,1]$, subject to the consistency condition
\beq
\lbl{consistency}
\frac{d\lambda}{d\tau}(1) = \Phi (1, \lambda (1), \lambda (0)),
\eeq
where the derivative in \re{consistency} is understood as a derivative
from the left.  
Such a specification of $\lambda$ on an interval of length 1
will be called a {\em generalized
initial condition}.

Then \re{delayeq} determines a unique solution
$\lambda (\tau)$ on some maximal interval $1 \leq \tau < \delta $
with $\delta \leq 2$.  
We shall show below that for the equations 
of interest to us, namely 
\re{rohreq2} and \re{acceq}, we actually have  $\delta = 2$, 
and $\lambda(\cdot)$
satisfies the equation on $[1,2]$, where 
$\lambda^\prime (2)$ is understood as a derivative from the left.
Iterating the process yields a solution $\lambda$ on $[0, \infty)$
whose values on $[0, 1]$ can be an arbitrarily specified $C^1$ function
satisfying the consistency condition \re{consistency}.

Iterating to the left to obtain a unique solution on $( - \infty, 0]$
determined by  $\lambda$ restricted to $[0,1]$
involves inverting $s \mapsto \Phi(\tau + 1, \lambda(\tau+1),  s)$ 
for fixed $\tau$.
For equations \re{rohreq2} or \re{acceq}, this is trivial.
For example for \re{rohreq2}, given $\theta(\tau)$ defined for
$0 \leq \tau < 1$, simply define $\theta$ on the ``preceding'' interval
$-1 \leq \tau < 0$
\beq
\lbl{leftiter}
\theta(\tau) := \sinh^{-1} (m\frac{d\theta}{d\tau} (\tau+1) -
E(\tau+1) ) + \theta(\tau+1),
\q -1 \leq \tau < 0,
\eeq
where the derivative $d\theta/d\tau (\tau+1)$ is understood as a derivative
from the right when $\tau+1 = 0$.
\section{Special cases of the general delay-differential equation
\re{delayeq}} 
\subsection{Mathematical preliminaries}
The delay-differential equation \re{delayeq} relates 
the solution $\lambda(\cdot)$ on an interval $[\alpha,\alpha + 1]$ to the 
solution on the ``preceding'' interval $[\alpha -1, \alpha]$.
The following proposition shows that for a class of equations
which includes \re{rohreq2} and \re{acceq} 
(the latter with $\theta$ regarded
as given, {\em a priori}),
the maximum of $\lambda$ on an interval $[\alpha, \alpha+1]$
is no greater than the maximum on the preceding interval 
$[\alpha-1, \alpha]$.  Similarly the minimum of $\lambda$ on 
$[\alpha, \alpha+1]$ is no less than the minimum on the preceding interval. 

For an arbitrary $C^1$ real-valued function $\lambda$ on the real line,
and arbitrary $\alpha < \beta$,  
define:
\begin{eqnarray}
\lbl{mdefs} 
M_{[\alpha, \beta]}(\lambda) &:=& 
\max \{\lambda(\tau)\  | \  \alpha \leq \tau \leq \beta \ \} \\ 
m_{[\alpha, \beta]}(\lambda) &:=& 
\min \{\lambda(\tau) \ | \  \alpha \leq \tau \leq \beta \ \} 
\end{eqnarray} 
\begin{proposition}
\lbl{prop1}
Let $ \tau, s \mapsto \Omega(\tau,s), \ -\infty < \tau, s, < \infty,$ 
be a $C^1$ function such that for each $\tau$, 
$s \mapsto \Omega(\tau, s)$ is a strictly increasing function satisfying 
$\Omega(\tau, 0) = 0 $.
Let $\lambda$ be a solution of a delay-differential equation of the  
special form
\beq
\lbl{specialeq}
\frac{d\lambda}{d\tau} = 
\Omega ( \tau, \lambda(\tau-1) - \lambda (\tau))
\q.
\eeq
Then 
for all $\alpha$, 
%for all $\alpha < \beta \leq \alpha + 1$,
\begin{eqnarray}
\lbl{prop1conc}
%M_{[\alpha, \beta]} &\leq& M_{[\alpha - 1, \alpha]} \\
%m_{[\alpha, \beta]} &\geq& m_{[\alpha - 1, \alpha]}  \q.  
M_{[\alpha, \alpha + 1]} &\leq& M_{[\alpha - 1, \alpha]} \\
m_{[\alpha, \alpha + 1]} &\geq& m_{[\alpha - 1, \alpha]}  \q.  
\end{eqnarray}
\end{proposition}
\begin{proof}
%It is sufficient to prove the case $\beta := \alpha + 1$.
For notational simplicity we take $\alpha := 0$.
Thus we will prove that
$$
M_{[1, 2 ]} \leq M_{[0, 1]}
\q.
$$ 
The proof of the corresponding assertion for $m$, which only requires
reversing a few inequalities, will be omitted.  
\newcommand{\taumax}{\tau_{\max}}

Let $\taumax $ denote a point in $[1,2]$ with
$$
\lambda (\taumax ) = M_{[1,2]}(\lambda)
\q.
$$
If $\taumax$ is an interior point of $[1,2]$,
then 
$$
0 = \lambda^\prime (\taumax) = 
\Omega(\taumax, \lambda (\taumax-1)- \lambda(\taumax))
\q,
$$
and since  $s \mapsto \Omega(\tau, s)$ is strictly increasing and
zero only at $s=0$, we have 
$$
M_{[0,1]} \geq \lambda(\taumax-1) = \lambda(\taumax) = M_{[1,2]}
\q.
$$

If $\taumax = 1$, then 
$$
M_{[0,1]} \geq \lambda(1) = \lambda(\taumax) = M_{[1,2]}
\q.
$$
If $\taumax = 2$, then we must have
\beq
\lbl{lambdadown}
\lambda^\prime (2) \geq 0 
\q,
\eeq
otherwise there would be points $\tau < 2$ with 
$\lambda(\tau) > \lambda (2)$ (i.e., the graph of $\lambda$ 
would be going down at 2),
contradicting $\lambda (2) = \lambda (\taumax) \geq \lambda (\tau)$
for all $1 \leq \tau \leq 2$.
Now from \re{lambdadown} and \re{specialeq},
$$
0 \leq \lambda^\prime (2) = \Omega (2, \lambda(1) - \lambda(2))
\q,
$$
so again using the fact that $\Omega(2, s)$ is increasing with $\Omega(2,0)=0$,
we conclude that $\lambda(1) - \lambda(2) \geq 0$. Finally,
$$
M_{[0,1]}(\lambda) \geq \lambda (1) \geq \lambda(2) = \lambda(\taumax) = 
M_{[1,2]}(\lambda).
$$ 
\end{proof}

Now we prove that given an arbitrary $C^1$ specification
of $\lambda(\cdot)$ on the interval $[0,1]$ 
satisfying the consistency condition \re{consistency}, 
there exists a solution $\lambda$ to \re{specialeq} 
defined on $[0, \infty)$ and taking the specified values on $[0,1]$.  
By a $C^1$ function on $[0,1]$, we mean a 
continuously differentiable function, with the derivatives at 0 and 1
understood as derivatives from the right and left, respectively.
\begin{proposition} 
\lbl{prop2}
Let $\tau \mapsto \psi(\tau)$ be a $C^1$ function on $[0,1]$ satisfying
the consistency condition \re{consistency}
Consider the equation \re{specialeq} of Proposition 1:
\beq
\lbl{specialeq2}
\frac{d\lambda}{d\tau} = 
\Omega ( \tau, \lambda(\tau-1) - \lambda (\tau)) 
\q, 
\eeq
where $\Omega$ satisfies the hypotheses of that Proposition.
Then there exists a unique $C^1$ solution $\lambda$ defined on 
$[0, \infty)$ and satisfying $\lambda(\tau) = \psi(\tau)$ 
for $0 \leq \tau \leq 1$.
\end{proposition}
\begin{proof}
If we regard $\lambda(\tau-1) = \psi (\tau-1)$ as given on $1 \leq \tau \leq 2$,
then equation \re{specialeq2} 
is covered by the standard existence and uniqueness 
theorems for ordinary differential equations (e.g., Perko, 1996, Chapter 2).
By these results, given the initial condition $\lambda(1)= \psi(1)$,
there exists a maximal interval $[1,\delta)$, $\ 1 < \delta \leq 2$,
such that there exists a $C^1$ solution $\lambda $ on $[1, \delta)$
satisfying that initial condition. 
Moreover, if $\lambda(\tau)$ remains in a compact set for 
$1 \leq \tau < \delta$, then $\delta = 2$ (Perko, 1996, Section 2.4, Theorem 2).

In other words, the only way the solution can fail to be globally
defined is if it blows up.  By Proposition \ref{prop1},
our solution does not blow up; hence it is defined for $0 \leq \tau \leq 2$. 
Iteration produces a solution defined on $[0, \infty)$. 
\end{proof}

We can also iterate to the left to obtain a solution defined on 
$(-\infty, \infty)$, assuming that for fixed $\tau$ 
we can invert $s \mapsto \Omega(\tau, s)$.  The inversion
is trivial for equations of the form $\re{rohreq2}$ and $\re{acceq}$,
so we have:
\begin{proposition}
\lbl{prop3} 
Let $\psi $ be a $C^1$ function on $[0,1]$ satisfying 
the consistency condition \re{consistency}.  
Then for any $C^1$ function $\tau \mapsto E(\tau)$,
there exists a $C^1$ solution $\tau \mapsto \theta(\tau)$
to the  DD1 equation \re{rohreq2}, 
defined for $-\infty < \tau < \infty$ and satisfying
$\theta (\tau) = \psi (\tau)$ for $0 \leq \tau \leq 1$. 
\end{proposition}
The same holds with $A$ in place of $\theta$ for the acceleration
equation \re{acceq}, provided we regard $\theta $ as given {\em a priori}
and assume that $E(\cdot)$ is $C^2$, so that $dE/d\tau$ is $C^1$. 

\subsection{Preacceleration and solutions runaway in the past}
The simple propositions of the previous subsection imply quite a lot.
Consider the DD1 equation \re{rohreq2}
$$
mA = m \frac{d\theta}{d\tau} = E + \sinh (\theta(\tau - 1 ) - \theta(\tau))
\q.
$$ 
for a continuous force $E$ applied for only a finite time, say for 
$0 \le \tau \le \tau_f $ (i.e., $E(\tau)$ vanishes off this interval).

Suppose that $A$ does not vanish identically on $[-1,0]$.
Then $A$ cannot vanish asymptotically in the past because
by Proposition \ref{prop1},
$$
M_{[-1,0]}(A) \leq M_{[-2, -1]}(A) \leq M_{[-3, -2]}(A) \leq \ldots 
\q,
$$
and 
$$
m_{[-1,0]}(A) \geq m_{[-2, -1]}(A) \geq m_{[-3, -2]}(A) \geq \ldots 
\q.
$$
The first equation tells us that 
if $M_{[-1, 0]}(A) > 0$, then $A$ assumes values at least 
as large as this infinitely often in the distant past. 
If, on the other hand, $M_{[-1, 0 ]}(A) \leq 0  $ , then 
also 
$m_{[-1, 0]}(A) \leq M_{[-1, 0]}(A) \leq 0 $, so $|A|$ assumes values at least 
at least as large as $|m_{[-1,0]}(A)|$ infinitely often in the distant past.  
So the only case in which $A$ could vanish asymptotically 
in the distant past is if 
$m_{[-1, 0]}(A) = M_{[-1, 0]}(A) = 0 $, but then $A$ vanishes identically
on $[-1,0]$. 

Thus the only physically reasonable initial specification of $A$
on $[-1, 0]$
is 
\beq
\lbl{initialacc}
A(\tau) \equiv 0 \q \mbox{for $ -1 \leq \tau \leq 0$}
\q.
\eeq
A glance at \re{acceq} shows that this implies that $A(\tau) $ vanishes
identically for $\tau \leq 0$, and so $\theta$ is constant (i.e., the velocity
is constant) for $\tau \leq 0$. 

So, we see that even for a compactly supported force (indeed, even
for identically zero force),
there {\em are} solutions of the DD equation 
which do not vanish asymptotically in the past.
These solutions also exhibit ``preacceleration''.
However, for a compactly supported force, 
we can choose the initial specification \re{initialacc} of $A$ 
so as to eliminate these pathological solutions. 

The situation for a force which is not compactly supported
seems unclear.  It does not seem obvious how to choose the initial
specification so as to force $A$ to vanish asymptotically 
as $\tau \arrow -\infty$ when $E$ does.
Since the theory is physically incomplete unless
one gives a prescription for an initial specification of $\theta$
on some interval of length 1, this is a point which advocates
of the DD equation should address.

Suppose we have agreed on such a prescription. 
We might define the $A$ determined by the prescription 
as {\em preaccelerative} if there exist two force
functions $E_1$ and  $E_2$ with $E_1 (\tau) = E_2 (\tau) $ 
for $\tau \leq 0$, but $A_1 (\tau) \neq A_2 (\tau)$
for some $\tau < 0$, where $A_1$ and $A_2$ denote the corresponding
accelerations obtained by solving the DD equation. 
It does not seem obvious that there is a prescription which will
outlaw preaccelerative solutions. 

\subsection{Postacceleration}
This section
considers a force applied for only a finite time $\tau_f$, 
say $0 \leq \tau \leq \tau_f$.   
For such a force, a {\em preaccelerative} solution of the DD equation
\re{rohreq} or  DD1 equation \re{rohreq2} will be  defined as one
such that the acceleration $A$ of the corresponding solution does not
vanish identically on $(-\infty, 0]$, 
and a {\em postaccelerative} solution is one
whose acceleration does not vanish identically on $[\tau_f, \infty)$.  

The discussion to follow explains
why either preacceleration or ``postacceleration'' 
is essentially built 
into the DD1 equation:  
one or the other must occur except in unusual special cases.  
This can be guessed from the observation that 
we have two ways to solve (uniquely) the 
DD1 equation \re{rohreq}: 
\beq
\lbl{rohreq2b}
mA = m \frac{d\theta}{d\tau} = E + \sinh (\theta(\tau - 1 ) - \theta(\tau))
\q.
\eeq
We may solve \re{rohreq2b} 
{\em forward in time} from a generalized initial condition 
$u(\tau) \equiv 0 $ for $-1 \leq \tau \leq 0$.
We can also solve 
{\em  backward in time} from a ``generalized final condition''
$u(\tau) \equiv 0$ for $\tau_f \leq \tau \leq \tau_f + 1$.
There is no reason to imagine that these two solutions will coincide,
and we shall now show that they almost never do.  

We may take $\tau_f = n$, with $n$ a positive integer.  
Consider the DD1 equation \re{rohreq2b} for $\tau_f = n \leq \tau \leq n+1$.  
Suppose there is no postacceleration, so that
$A = d\theta/d\tau$ vanishes on this interval. 
Then $\theta$ is constant on this interval, say
$\theta(\tau)= k $,  a constant,  for 
 $\tau_f = n \leq \tau \leq n+1$.  
Substituting these facts  into \re{rohreq2b}
gives 
$$0  = 0  +  \sinh (\theta(\tau - 1) - k) \q \mbox{for}\q 
n \leq \tau \leq n+1 \q,
$$
which may be restated as:
$$
\theta (\tau) \equiv k \q\mbox{for}\q n-1 \leq \tau \leq n \q.  
$$

Next consider equation \re{rohreq2b} on the 
interval  $n-1 \leq \tau \leq n$.
Using the above observation that $\theta(\tau) \equiv k$ on this interval, 
which implies that $A \equiv 0$ on this interval, we see that
\beq
\lbl{lastint}
E(\tau)  = - \sinh (\theta(\tau - 1) - k) \q \mbox{for}\q 
n-1 \leq \tau \leq n \q.
\eeq
This says that, {\em given} the values of $\theta (\tau) $ 
for $n-2 \leq  \tau \leq n-1$ (i.e., the values of 
$\theta(\tau - 1)$ for $n-1 \leq \tau \leq n$), 
there is a {\em unique} force function $E(\tau)$
on $n-1 \leq \tau \leq n$ which will eliminate postacceleration.
Put differently, for any {\em other} force function
on $n-1 \leq \tau \leq n$, postacceleration 
{\em must} occur.

To complete the argument, suppose there is no preacceleration,
i.e., $A(\tau) \equiv 0$ for $\tau < 0$.
Then for some constant $\theta_0$, $\theta(\tau) = \theta_0$ for $\tau < 0$.
Solve the DD1 equation \re{rohreq2}
forward in time starting with this generalized initial condition,
$\theta(\tau) \equiv \theta_0$ for $-1 \leq \tau \leq 0$.
This will produce a unique solution $\theta(\tau)$ on the interval
$0 \leq \tau \leq n-1$, and this solution depends only on the values
of $E(\tau)$ for $0 \leq \tau \leq n-1$.
Substituting this solution in equation \re{lastint} 
shows that 
postacceleration must occur for any force function
$E(\tau)$ on $ n-1 \leq \tau \leq n$ which does {\em not} satisfy
equation \re{lastint}, i.e., for almost all force functions.

The above argument becomes much simpler for the case $n=1$,
corresponding to a force applied only for a time interval of length 1,
i.e., a time interval the length of the delay parameter in the DD equation. 
In Rohrlich's motivation of the DD equation, this time interval
is the time light required to traverse the spherical particle's diameter,
so it is probably too small to have any observational interest, 
but an examination of this special case is a quick way for the
reader to become convinced that it is mathematically unreasonable
to expect to simultaneously eliminate both preacceleration and
postacceleration in general.

The simpler argument for $n=1$ goes as follows.  Observe as before that
$\theta(\tau)$ must be constant for $n-1 \leq \theta \leq  n$, i.e.,
$\theta (\tau) = k$ for   $0 \leq \theta \leq 1$,  
and in particular, $\theta(0) = k$.
If there is no preacceleration, then $\theta(\tau)$ must be constant
for $\tau < 0$, say 
$\theta(\tau) \equiv \theta_0$, for $\tau \leq 0$. In particular,
$\theta_0 = \theta(0) = k$.
Hence on $[0,1]$, 
$$
 \theta(\tau - 1) = \theta_0 = k  \q\mbox{for}\q 0 \leq \tau \leq 1 \q.
$$
Substitute this in \re{lastint} to obtain
$$
E(\tau) = -\sinh(k - k)  \equiv 0
\q\mbox{for} \q 0 \leq \tau \leq 1\q.
$$ 
This says that for an force applied no longer than the time
delay, either precceleration or postacceleration {\em must} occur
except in the trivial case of identically  vanishing force. %
\section{Forward asymptotics for eventually vanishing force} 
This section proves 
that solutions of the 
DD1 equation \re{rohreq2} for eventually vanishing force have 
proper acceleration asymptotic to zero and asymptotically constant velocity  
in the future:
\beq
\lbl{asymptotics}
\lim_{\tau \goesto \infty} A(\tau) = 0 \q; \q 
\lim_{\tau \goesto \infty} \theta(\tau) = \mbox{constant} 
\q.
\eeq 
Thus replacing the Lorentz-Dirac equation by the DD equation 
does eliminate the undesirable ``runaway'' solutions of the former
for the special case of motion in one space dimension.
Whether this is also true for motion in three space dimensions
is unknown.

Suppose that $E $ eventually vanishes, and
choose the origin of time so that $E(\tau) = 0$ for $\tau \geq 0$.
Then for $0 \leq \tau \leq 1$, $A$ satisfies
\beq
\lbl{secFAeq1}
m \frac{dA}{d\tau} = 
\cosh (\theta (\tau-1) - \theta(\tau)) (A(\tau-1) - A(\tau))
\q
\eeq 
We shall consider the related equation
\beq
\lbl{heq}
 \frac{dh}{d\tau} = \phi (\tau) (g(\tau) - h(\tau)),\ 
0\leq \tau \leq 1,\  \mbox{with initial condition $h(0) = g(1)$}. 
\eeq
Here $\phi$ will be a given continuous function on $[0,1]$,
and $g$ another function with the same domain.
We will think of $\phi$ as fixed until further notice,
and of \re{heq} as defining a mapping which assigns to each $g$ another 
function $h$, namely the unique solution of \re{heq}. 
We will show that this mapping is a strict contraction
relative to a certain Banach space norm. 

Let $C[0,1]$ denote the real Banach space of all continuous real-valued
functions on $[0,1]$ with the supremum norm $\| \cdot \|_\infty$:
for $g \in C[0,1]$, 
$$
\| g \|_\infty := \max\{\, |g(\tau) |\ | \ 0 \leq \tau \leq 1\}
\q.
$$
For $g \in C[0,1]$, let $M(g) := M_{[0,1]}(g)$ and 
$m(g) := m_{[0,1]} (g)$ denote the maximum and minimum
of $g$.

Let $B$ denote the Banach space which is the quotient of $C[0,1]$
by the one-dimensional subspace of constant functions.
For $g \in C[0,1]$, let $\tilde{g}$ denote its image in $B$ under the 
quotient map.
\newcommand{\one}{{\bf 1}}
Denoting by $\one$ the function constantly equal to 1,
and by 
\newcommand{\bbR}{{\bf R}}
$\bbR$ the real field,
the quotient norm $\|\cdot\|$ on $B$ is:
\beq
\lbl{quotnorm}
\| \tilde{g} \| := \inf \{ \, \| g + \alpha \one \|_\infty \ |\  \alpha \in \bbR\ \}
= \frac{1}{2} [ M(g) - m(g)]
\q.
\eeq 
The first equality is the definition of the quotient norm,
and the last is a simple exercise.

Let $\phi$ be a given function, and consider the linear mapping
$Q_\phi : C[0,1] \rightarrow C[0,1]$ defined as follows. 
For any $g \in C[0,1]$, 
\beq
\lbl{Qdef}
Q_\phi g := h
\q,
\eeq
where $h$ is the unique solution of \re{heq}. 
%In contexts where the dependence of $Q$ on $\phi$ is unimportant,
%we sometimes write $Q$ in place of $Q_\phi$,
%and we also write $Q_\phi g$ or $Qg$  
%in place of $Q_\phi (g)$.
%
\newcommand{\tQ}{\tilde{Q}}

Define $\tQ_\phi: B \arrow B$ to be the analog of $Q_\phi$ on the quotient space
$B$:  for all $g \in C[0,1]$, 
\beq
\lbl{tQdef}
\tQ_\phi (\tilde{g}) := \widetilde{Q_\phi g}
\eeq
We see that $\tQ_\phi$ is well-defined by noting that for any $\alpha \in \bbR$,
$$
\frac{d(h + \alpha \one)}{d\tau} = \frac{dh}{d\tau} 
= \phi(\tau)[g(\tau) - h(\tau) ] 
= \phi(\tau)[(g(\tau) + \alpha) - (h(\tau) + \alpha)] 
\q, 
$$
and also $(h + \alpha \one)(0) =h(0) + \alpha = g(1) + \alpha 
= (g+\alpha\one)(1).$ 
Thus if we alter $g$ by adding a constant $\alpha$ to it,
then we also alter the solution of \re{heq} by 
the same additive constant, so 
that the $h$ in \re{tQdef} depends only on the equivalence class of $g$ in
$C[0,1]$ modulo the constants.

Next we show that for any non-negative $\phi$, 
the mapping $\tQ_\phi : B \arrow B $ is a strict contraction.  
Actually, we'll need the following stronger fact 
giving a uniform bound on  $\| \tQ_\phi \|$ 
for any bounded set of non-negative $\phi$: 
\begin{lemma}
\lbl{lemma4}
For each non-negative $\phi \in C[0,1]$, $\| \tQ_\phi \| < 1$.
Moreover,
given any constant $k_0$,
there is a constant $ k< 1$ such that  
for all non-negative functions $\phi$ with $ \|\phi\| \leq k_0 $, 
and for all $g \in C[0,1]$,
\beq
\lbl{contraction}
\| \tQ_\phi \tilde{g} \| \leq k \|\tilde{g}\|
\q.
\eeq
\end{lemma}
\begin{proof} 
For each non-negative function $\phi \in C[0,1]$, define 
$$
\psi(x) := \int_0^\tau \phi(s) \, ds
\q.
$$ 
Note that since $\phi$ is non-negative, $\psi$ is non-decreasing. 

Then for any $g \in C[0,1]$,   
$ (Q_\phi g)(\tau) = h(\tau)$
with $h$ the solution to  \re{heq}:
$$  
h(\tau) = 
e^{-\psi(\tau)}g(1) + 
e^{-\psi(\tau)} \int_0^\tau e^{\psi(s)}\phi (s) g(s) \, ds
\q,
$$
so
\begin{eqnarray*}
\lbl{lemma4eq1}
h(\tau) &\leq&  
e^{-\psi(\tau)}g(1) + 
M(g) e^{-\psi(\tau)} \int_0^\tau e^{\psi(s)} \frac{d\psi(s)}{ds} \, ds \\
&=& 
e^{-\psi(\tau)}g(1) + 
M(g) (1 - e^{-\psi(\tau)}) \\
&=& M(g) + (g(1) - M(g))e^{-\psi(\tau)}
\q.
\end{eqnarray*}
Since $g(1)-M(g) \leq 0$ and $\psi$ is non-decreasing, 
\begin{eqnarray}
\lbl{lemma4eq2}
M(h) &\leq& M(g) + [g(1) - M(g)] e^{-\psi(1)} \nonumber\\
&=& M(g)(1 - e^{\psi(1)}) + g(1)e^{-\psi(1)}
\q.
\end{eqnarray}
Similarly,   
$$
h(\tau) \geq 
e^{-\psi(\tau)}g(1) + 
m(g) (1 - e^{-\psi(\tau)})
\q,
$$
and since $g(1) - m(g) \geq 0$ and $\psi$ is non-decreasing, 
\begin{eqnarray}
\lbl{lemma4eq3} 
m(h) &\geq& m(g) + [g(1) - m(g)]e^{-\psi(1)} \nonumber\\
&=& m(g)(1-e^{-\psi(1)}) + g(1)e^{-\psi(1)}	
\q.
\end{eqnarray}
Subtracting the inequalities \re{lemma4eq2} and \re{lemma4eq3} 
shows that for each non-negative $\phi$, 
$\tilde{Q_\phi}$ is a strict contraction:
\begin{eqnarray}
\lbl{lemma4eq4}
\| \tQ_\phi (\tilde{g}) \| &=& [M(h) - m(h)]/2 \non \\ 
&\leq& (1 - e^{-\psi(1)})[M(g)-m(g)]/2 \ =\  (1-e^{-\psi(1)})\|\tilde{g}\| 
.
\end{eqnarray}
Finally, the uniformity condition follows from \re{lemma4eq4}
with $k := 1 - e^{-k_0}$, since $\| \phi \| \leq k_0$ implies
$\psi(1) \leq k_0$ and hence $ 1-e^{-\psi(1)} \leq 1-e^{-k_0}$.
\end{proof}
Now we prove the main result of this section:
\begin{theorem}
\lbl{thm5}
Suppose that the force $\tau \mapsto E(\tau)$ in the DD1 equation 
\re{rohreq2} eventually vanishes, meaning that there exists a
proper time $\tau_0$ such that $E(\tau) = 0 $ for $\tau \geq \tau_0$.
Then there exists a constant $\theta_\infty$
such that 
\beq
\lbl{thm5conc}
\lim_{\tau \goesto \infty} A(\tau) = 0 \q \mbox{and} \q 
\lim_{\tau \goesto \infty} \theta(\tau) = \theta_\infty 
\q.
\eeq
\end{theorem}

\begin{proof}
We choose the origin of proper time so that $E(\tau) = 0$ for
$\tau \geq 0$. 
In solving the acceleration equation \re{acceq},
we assume that the solution $A$ is already obtained for $\tau \leq 0$.
Then the solution on $[0,1]$ is obtained by applying an opertor $Q_\phi$ to 
the restriction $A | [-1,0]$ of $A$ to $[-1,0]$, translated one unit right. 

More precisely, define an operator $T$ on $C[0,1]$ to be right translation
by one unit:  $(Tf)(\tau) := f(\tau - 1)$ for $f \in C[0,1]$ .
Let 
$\phi (\tau) := \cosh(\theta(\tau-1) - \theta(\tau))$, $\ 0\leq \tau\leq 1$. 
Then 
$$
\mbox{for $0\leq \tau \leq 1$}, \q 
A(\tau) = Q_\phi T(A|[-1,0])(\tau)
\q.
$$

Though we don't know $\theta(\tau)$ for $0 \leq \tau \leq 1$,
we do know from Proposition 1
that it is bounded above and below by bounds no worse
than on the preceding interval $-1\leq \tau\leq 0$.
Hence we have the {\em a priori} bound 
$$\| \phi \|_\infty \leq  
k_0 := \cosh(|M(\theta|[-1,0])| + | m(\theta|[-1,0]) |)
\q.
$$ 

The solution $A(\tau)$ for $1 \leq \tau \leq 2$ is similarly obtained,
except that the $Q_\phi$ is different because the $\phi$ is different.  
However, Proposition \ref{prop1} shows 
that we have the same {\em a priori} bound on the new $\| \phi \|_\infty$. 
After $n$ applications of Lemma \ref{lemma4}, 
we find that
$$ 
(M(A|[n-1,n]) - m(A|[n-1, n]))/2
= \| A|[n-1,n] \| \leq  k^n \| A|[-1,0] \| \q.
$$ 
This implies that 
$A|[n-1,n]$, 
becomes asymptotically constant as $n$ becomes large.
The constant must be zero because $A = d\theta /d\tau$, and the previous section
showed that the rapidity $\theta$ is bounded
on $[0,\infty]$.  Hence 
\beq
\lbl{accgoesto0}
\lim_{\tau\goesto\infty} A(\tau) = 0 
\q.
\eeq	

Finally, we show that the rapidity (hence the velocity) 
becomes asymptotically constant in the future:  
for some constant $\theta_\infty$,
\beq
\lbl{rapgoesto0}
\lim_{\tau\goesto\infty} \theta(\tau) = \theta_\infty
\q.
\eeq 
From the last section, $M_{[n,n+1]} (\theta)$ is a nonincreasing 
function of the positive integer $n$, and $m_{[n,n+1]}(\theta)$
is nondecreasing.  This implies that if \re{rapgoesto0} were false,
there would exist a positive constant $\delta$ with
$M_{[n,n+1]}(\theta) - m_{[n,n+1]}(\theta) \geq \delta$ 
for all $n \geq 0$.  Choose points $\tau^+_n$ and $\tau^-_n$
in $[n, n+1]$ with 
$\theta(\tau^+_n) = M_{[n,n+1]}(\theta)$
and
$\theta(\tau^-_n) = m_{[n,n+1]}(\theta)$.
Suppose $\tau^-_n \leq \tau^+_n$.
Applying the Mean Value Theorem to the interval $[\tau^-_n, \tau^+_n]$ 
yields a point $\tau_n$ between $\tau^-_n$ and $\tau^+_n$ with 
\beq
\lbl{Anot0}
A(\tau_n) = \frac{\theta(\tau^+_n) - \theta(\tau^-_n)}{\tau^+_n - \tau^-_n}
\geq \frac{M_{[n,n+1]}(\theta) - m_{[n,n+1]} (\theta)}{1} \geq \delta
\q.
\eeq
Similarly, if $\tau^+_n < \tau^-_n$, then $A(\tau_n) \leq -\delta $ 
for some $\tau_n \in [n,n+1]$; both this and \re{Anot0} contradict  
the previously established fact that $\lim_{\tau \goesto \infty} A(\tau) = 0$. 
\end{proof}
\section{Summary and assessment}
The above proofs were given in the context of the normalized 
DD equation \re{rohreq} for motion in one space dimension
and its equivalent formulation, the DD1 equation \re{rohreq2}.
Translating them  into information about the original 
DD equation \re{rohreq1} discussed in (Rohrlich, 1997, 1999) 
yields the following observations, in which the motion is assumed
restricted to one spatial dimension unless otherwise specified.
\begin{enumerate}
\item
An appropriate condition to guarantee a unique solution 
is a specification of the rapidity $\theta(\tau)$ on
some closed interval $[\alpha, \alpha + \tau_1]$,
subject to the consistency condition \re{consistency} (written there for
$\alpha := 0$ and $\tau_1 = 1$). 
We refer to this specification as a ``generalized initial condition''.
\s
There exists a unique $C^1$ solution $\theta(\cdot)$ of \re{rohreq2} satisfying any 
$C^1$ generalized initial condition, and hence a unique four-velocity
$u(\tau) = (\cosh \theta(\tau), \sinh \theta(\tau) )$ and 
worldline $\tau \mapsto z(\tau)$ with $dz/d\tau = u(\tau)$
(by integration, with arbitrary specification of $z(0)$). 
\item
For the case of a force $\tau \mapsto E(\tau)$ 
applied for only a finite time
(i.e., compactly supported force), 
there exist solutions which do not vanish asymptotically 
in the distant past.  
Such solutions seem unphysical.
For simplicity of language,
we will call them ``past runaway'' solutions even though we do not know
that the acceleration becomes unbounded in the infinite past.
\s
However, for compactly supported force,
there exists a unique choice of generalized initial condition
which eliminates these past runaway solutions.
This seems the physically relevant choice of initial condition.
\item
For a force not compactly supported, 
it has not been proved that
there is an appropriate choice of generalized initial condition 
to eliminate past runaways.
Indeed, it seems unclear what such a choice might be.  
If the DD equation is to be used to predict the motion of actual
particles, some definite procedure for specifying 
the generalized initial condition would be necessary. 
\item
For a force $E$ which eventually vanishes (i.e.,  for some $\tau_0$,
$E(\tau) = 0$ for $\tau \geq \tau_0$),  
the acceleration does converge to zero in the future, and the 
velocity becomes asymptotically constant. 
Thus for the DD equation for motion in one space dimension, 
analogs of the pathological 
``future runaway'' solutions of the Lorentz-Dirac equation do not occur.
It is unknown whether future runaways are also impossible 
for general three-dimensional motion obeying the DD equation. 
\item
For compactly supported force, there exist ``preaccelerative'' solutions
(defined as solutions with nonzero acceleration {\em before}
the force is applied).  
Such solutions seem unphysical because they violate ``causality''. 
However, the same generalized initial condition which rules out past
runaways also rules out these undesirable preaccelerative solutions. 
\item
For compactly supported nonzero force and generalized initial condition
chosen to rule out preacceleration, 
the resulting solution necessarily exhibits 
the unusual property which we call
``postacceleration'':  
the acceleration persists after the force is removed, 
and in fact into the infinite future.
Some may consider this strange behavior sufficiently unphysical  
to rule out the DD equation. Others may welcome the unusual prediction
as a potential physical test which if verified, 
would constitute striking evidence for the equation. 
\end{enumerate}

Given that known motivations of the DD equation are not fundamental
(e.g., employ uncontrolled approximations 
obtained by ignoring nonlinear terms), and given the 
uncertainties noted above about the mathematical appropriateness of 
the equation, it seems that more work would be needed to convert
it from a speculative proposal to an accepted physical principle. 

%Definitive experimental evidence seems out of reach for all proposed
%classical equations of motion for charged particles.
%Absent such evidence, many of the disputes in this area seem
%more theological than scientific.

It is often said that a resolution of the logical problems of classical
electrodynamics must come from more fundamental physical principles
of quantum mechanics.  This may indeed turn out to be true, 
but it is not the only possibility.
It would seem strange if a logically consistent and physically sensible
classical electrodynamics were inherently impossible.
Although nearly all of physics rests in principle on quantum mechanics,
there do exist consistent and sensible classical theories in many areas.
The present theory of classical charged particles cannot be considered
sensible.

It is interesting 
that the Lorentz-Dirac equation has survived as the principal candidate for 
a classical equation of motion despite predictions so bizarre
(e.g., Eliezer, 1943, Parrott, 2002) that no one will admit to believing them. 
The reason for the survival may be the fundamental nature of
Dirac's derivation of the equation.  If one accepts mass renormalization
(admittedly controversial),
then one can convincingly obtain the Lorentz-Dirac
equation from the principle of conservation of energy-momentum 
with no approximations whatever.   
One does not lightly discard such mathematically tight arguments.  

	The motivations of many other proposed equations of motion 
are aesthetically less pleasing.    
Most consist of more or less 
{\em ad hoc} modifications of the Lorentz-Dirac equation
which do not {\em obviously} lead to its bizarre predictions.
(This is not the same as saying that the modifications obviously {\em do not}
lead to similarly bizarre predictions!) 

Assuming that one is willing to believe in postacceleration,
the DD equation seems physically possible,
but not physically compelling.
It is not clear in what sense, if any,
solutions of the DD equation conserve energy-momentum.  
It would be desirable to find
a convincing, mathematically rigorous way to relate the DD equation
to the principle of conservation of energy-momentum.  
\section{Appendix 1}
When the body of this paper was written, 
I was unaware of seminal work of Ryabov (Ryabov,
1960 , 1961, 1963, 1965)
on delay differential equations,  
subsequently refined and extended by (Driver, 1968).  
This work does not directly
apply to the DD equation because this equation does not satisfy
Driver's hypotheses.  However, it has the same structure as equations
considered by Driver.  

If $\sinh $ were a bounded function,
then Driver's results {\em would} apply to
Rohrlich's equation  \re{rohreq2}.
Thus we can hope that conclusions similar to Driver's might apply
to the DD equation. 
If so, they would shed light on  
some of the problems mentioned in the Summary and Assessment section.

This appendix outlines Driver's conclusions in the context of 
the DD equation.
It is mathematically based on the excellent 
semi-expository paper (Driver, 1968), but the presentation
follows the introductory sections of (Chicone, 2003),
which is based on Driver's work.
The latter presents a point of view 
particularly congenial to a discussion of the DD equation. 

We shall be concerned with delay-differential equations of 
the form 
(\thedelayeqnum ), but for easy exposition we need to
reinsert the positive delay parameter $\tau_1$ 
(which was set to 1 in (\thedelayeqnum) by appropriate choice of units): 
\beq
\lbl{delayeq2}
\frac{d\lambda}{d\tau} = \Phi ( \tau, \lambda (\tau), \lambda (\tau-\tau_1))
\q.
\eeq
In equation (\thedelayeqnum), the function $\lambda$ was a real-valued
function, but the case in which $\lambda$ takes values in $R^n$
is no more difficult to discuss, and the results that we shall present
apply also to this case.  
Therefore, we assume that $\lambda(\tau) \in R^n$.

The specialization of Driver's work of interest to us applies 
to equations of this form with $\Phi$ continuous and bounded.  
The hypothesis that $\Phi $ be continuous and bounded 
makes the results easy to state.
It can be relaxed, but not so far as to admit the Rohrlich
equation.   More precisely, 
the hypothesis that $\Phi $ be bounded can be relaxed 
to an assumption that $\Phi$ satisfy a certain Lipschitz condition.

A very important hypothesis for the results of Ryabov and Driver
to be described is that $\tau_1$ be sufficiently small.
That is why we needed to reinsert the delay as an explicit parameter
in \re{delayeq2}.

Previously we normalized
$\tau_1$  to 1 by appropriate choice of units,
and the reader may be puzzled as to why we can't  
maintain the same normalization here. 
We could, but it's not convenient.
The reason is that normalizing $\tau_1$ to 1 
correspondingly changes the Lipschitz constant, so that 
the restriction that $\tau_1$ be sufficiently small is replaced 
in general by a restriction on the Lipshitz constant, or by a restriction
on the bound of $\Phi$ under our simpler hypothesis that $\Phi$
be continuous and bounded.
From a physical point of view, it is more natural to phrase the hypothesis
for the Ryabov/Driver results in terms of a restriction on the time delay 
instead of a restriction on the Lipshitz condition or the bound of $\Phi$. 

A priori, a mathematically  appropriate initial condition  
to specify a unique solution for \re{delayeq2}
is a specification of the function $\lambda(\cdot)$ 
on an interval $[0, \tau_1)$ subject to the consistency condition
$d\lambda/dt(\tau_1) = \Phi(0,\lambda(\tau_1), \lambda(0))$. 
Thus the solution space of \re{delayeq2} is infinite dimensional.

For the DD1 equation \re{rohreq2} 
(corresponding to $n=1$ in \re{delayeq2}), 
the physical expectation is that there should be precisely one 
physically relevant solution for each ordinary initial condition
$\lambda(0) = \lambda_0 \in R^1$.
Thus the space of mathematical solutions of \re{delayeq2}
is too large---one needs to find a condition which will reduce 
the infinite-dimensional solution space to a one-dimensional space.
The Ryabov/Driver theory suggests such a condition,
as we shall now discuss.  

A solution $\lambda$ of \re{delayeq2} is called a {\em special solution}
if 
\beq
\lbl{eqa2}
\sup_{-\infty < \tau < \infty}
e^{- |\tau|/\tau_1} \| \lambda(\tau) \| <\infty \q.
\eeq
Thus a special solution is one which does not increase at too fast 
an exponential rate. In particular, all solutions which are 
{\em not} special are ``runaway'' (either in the past or the future), 
and therefore presumably unphysical for physically possible forces. 

Driver's results imply that for every $\lambda_0 \in R^n$,
there exists a unique special solution 
(to \re{delayeq2} with $\Phi$ continuous and bounded)
satisfying $\lambda(0) = \lambda_0$.  
Thus the space of special solutions is a manifold of
dimension $n$, coordinatized by the map $\lambda \mapsto \lambda(0)$.

In addition, Driver showed that this manifold is an attractor
in the sense that {\em every} solution approaches (is ``attracted'' to)
a special  solution (exponentially fast) in the infinite future. 
Thus the manifold of special solutions models the long-term behavior
of all solutions.  

The DD equation \re{rohreq} in 
$m$-dimensional space ($(m+1)$-dimensional spacetime),
can be reformulated as an equivalent equation of 
form \re{delayeq2} with $n = m$.  
Here 
we are primarily interested in the case $m=3$, and secondarily  
in $m=1$.
The reason that \re{rohreq} with $m=3$ is not immediately 
of the form \re{delayeq2}
is that a four-velocity $u$ must satisfy 
the auxiliary condition $u^\alpha u_\alpha = 1$. 
Taking this into account 
reduces it to an equivalent equation of form \re{delayeq2} with $m = 3$;
we omit the details.
Although the corresponding $\Phi$ on its right side is not bounded,
we can hope that the conclusions of the Ryabov/Driver theory might
still apply.

Suppose that the force in the DD equation \re{rohreq} is  
continuous and compactly supported (hence bounded),
and consider the equivalent equation of form \re{delayeq2}
with $m = 3$.  
Then all physically relevant solutions
should be bounded, hence special (because all non-special solutions
are ``runaway'', as noted above).
If the conclusions of the Ryabov-Driver theory extend to this more general
situation, then the space of physically relevant solutions has 
dimension no greater than $m$.   
 
In general, special solutions can increase exponentially fast, 
so we cannot immediately identify special solutions
with ``physical'' solutions.
However, it is unknown whether exponential growth
of special solutions can occur for the DD equation
in four-dimensional spacetime.  

If there were exponentially increasing special solutions to the DD equation
in four-dimensional spacetime with compactly supported force, 
this would seem to definitively rule
it out as a physically realistic equation of motion. 
However, it is an attractive conjecture that in this situation,
all special solutions must be bounded. 
If so, then the space of physically relevant solutions
coincides with the space of special solutions,
and it has the physically expected dimension 3.
Similar remarks apply to the DD equation 
in spacetimes of other dimensions.

More generally, it is attractive to speculate 
that under reasonable hypotheses on a force not compactly supported
(e.g., an exponentially decreasing force), 
all special solutions might be bounded.  
If this could be proved, it might be considered as a solution
in principle to the problem posed in comment 3 of the Summary and Assessment
section:  Is there a generalized initial condition which 
will eliminate runaways in this situation? 
That is, we could identify ``physical'' (e.g., non-runaway) solutions
with special solutions.
However, since  closed  form special solutions  are typically
hard to come by,  the practical problem of identifying the special solution
in some constructive way would remain.

The only thing which prevents us from directly applying the insights
of the Ryabov/Driver theory to the DD1 equation
\re{rohreq2} is that 
the function $\sinh$ does not satisfy the required Lipschitz condition.
But it seems possible that this may be merely a technical matter,
and it seems reasonable to hope that the Ryabov/Driver theory might be 
extended to apply to this equation. 
If so,
Theorem 5's conclusion 
that the rapidity $\theta(\tau)$ converges 
to a constant as $\tau \rightarrow \infty$ illustrates 
how the manifold of special solutions models the long-term behavior
of the system.  
If every special solution converges to a constant
in the future (as the theorem establishes for eventually vanishing force), 
then {\em every} solution 
must similarly converge to a constant (as the theorem also establishes). 

However, even if the Ryabov/Driver theory could be extended 
to cover the DD equation \re{rohreq},
there would be another difficulty
which could be more fundamental.
This difficulty is that even for delay equations \re{delayeq2}
to which the Ryabov/Driver theory does apply, it only applies
for sufficiently small $\tau_1$.%
\footnote{See, e.g., (Chicone, 2003) for 
an explicit bound $\tau_1 < \delta$ which guarantees applicability
of the Ryabov/Driver results.  Unfortunately for our application, 
the bound $\delta$ is inversely proportional to the Lipschitz constant, 
which is formally infinite for $\sinh$ (so that $\delta$ is formally zero). 
}

Unfortunately, in our application, it is questionable to assume
that $\tau_1$ is arbitrarily small.
Indeed, a critic of an earlier version of 
the present work raised the following interesting
objection.

In Rohrlich's formulation, the classical charged particle is treated
as a sphere of nonzero radius and the delay $\tau_1$ is 
twice that radius (the time for light to traverse the sphere).
If the radius is too small, then the classical electrostatic 
energy of such a sphere will exceed the typical energy required
for the quantum effect of pair production, moving the problem
out of the domain of classical electrodynamics
and into the domain of quantum electrodynamics. 
But the DD equation is proposed solely as a classical equation of motion.

Put differently, the considerations leading to the DD equation 
are not claimed to apply to arbitrarily  small delays.  
The critic mistakenly thought that the present work assumed 
Caldirola's value of 
$\tau_1 = 4q^2/(3mc^3) \approx 1.2 \times 10^{-23} \mbox{sec}$,
which he believed would be far less than could be the radius of
any classical charged particle. 

Therefore,
any proposed application of the Ryabov/Driver theory 
to the DD equation within the framework of motivation 
via a nonzero particle radius 
should address the question of whether 
the delay (equivalently, the particle's radius) 
is small enough for an extended Ryabov/Driver theory to apply. 
\section{Appendix 2} 
This appendix gives the details of the equivalence of 
\newcounter{saveequation}
\setcounter{saveequation}{\value{equation}}
\setcounter{equation}{\therohreqonenum}
\addtocounter{equation}{-1}
\beq 
m_1 \frac{du^i}{d\tau} = 
f^i(\tau) +  
m_2 [ u^i(\tau-\tau_1) - u^\alpha (\tau - \tau_1) u_\alpha (\tau) u^i(\tau) ] 
\q,
\eeq
and 
\newcommand{\etabar}{\bar{\eta}} 
\newcommand{\Abar}{\bar{A}} 
\newcommand{\zbar}{\bar{z}} 
\setcounter{equation}{\therohreqbarnum}
\addtocounter{equation}{-1} 
\beq
 m \frac{d\ubar ^i}{d\taubar} = \fbar ^i + 
 [ \ubar^i(\taubar-1) - \ubar^\alpha (\taubar - 1) 
\ubar_\alpha (\taubar) \ubar^i(\taubar) ] 
\q.
\eeq
We'll show that these equations are equivalent if the barred quantities 
are defined by:
\setcounter{equation}{\value{saveequation}}
\begin{eqnarray}
\lbl{ubardef}
\ubar (s) &:=& \tau_1 u(\tau_1 s) \q \mbox{for all real numbers $s$,}\\
\lbl{fbar}
\fbar (s) &:=& \tau_1 f(\tau_1 s)/m_2 \q 
\mbox{for all real numbers $s$, and}\\
\lbl{m}
m &:=& m_1/(m_2\tau_1)\q.
\end{eqnarray} 

Call the time unit with respect to which \re{rohreq1} is written the 
``old time unit''.
First we change to a new time unit which equals $\tau_1$ old time units. 
If the old metric tensor is denoted $\eta := \mbox{diag}(1,-1,-1,-1)$,
this can be accomplished by introducing 
a new metric $\etabar := \tau_1^{-2}\eta$.
This changes the old proper time $\tau$ to a new proper time
$$\taubar = \tau_1^{-1} \tau \q,\q \tau = \tau_1\taubar\q.$$

A worldline $\tau \mapsto z(\tau) \in R^4 $ 
parametrized by old proper time $\tau$
corresponds to a ``new'' worldline
$\taubar \mapsto \zbar(\taubar)$ parametrized by new proper time $\taubar$,
with $\zbar(\taubar) = z(\tau)$, i.e., for any real number $s$,
$$
\zbar(s) = z(\tau_1 s) 
\q.
$$
The ``new'' worldline consists of the same set of points as the old
worldline, but the parametrization is different.  

The four-velocity $\ubar(\taubar) $ of the new worldline 
is related to the old four-velocity $u(\tau)$ by:
$$
\ubar(\taubar) := \frac{d\zbar}{d\taubar} = \tau_1 u(\tau),\q
\mbox{equivalently,}\q  u(\tau) = \tau^{-1}_1 \ubar(\taubar)
$$
Thus $\ubar$ and $u$ are related by the simple transformations
\beq
\lbl{ubarandu}
\ubar(s) = \tau_1 u(\tau_1 s), 
\q\mbox{and}\q
u(s) = \tau^{-1}_1 \ubar(\tau^{-1}_1 s), \q \mbox{for all real numbers $s$}.
\eeq
This gives a way to translate any statement about $\ubar$ into
a corresponding statement about $u$, and vice versa.
Similarly,
\beq
\lbl{abar}
\frac{d\ubar}{d\taubar} = {\tau_1}^2 \frac{du}{d\tau}, 
\q\mbox{equivalently,}\q  
\frac{du}{d\tau} = \tau^{-2}_1 \frac{d\ubar}{d\taubar} . 
\eeq

Next, divide \re{rohreq1} by $m_2$, and substitute the above relations
\re{ubarandu} and \re{abar}, obtaining 
\beq
\lbl{trans}
\frac{m_1}{m_2} \tau^{-2}_1 \frac{d\ubar^i}{d\taubar} = 
\frac{f^i(\tau)}{m_2} +  
{\tau_1}^{-1} 
[ \ubar^i(\taubar-1) - \ubar^\alpha (\taubar - 1) 
\ubar_\alpha (\taubar) \ubar^i(\taubar) ] 
\q.
\eeq 
Now inspection shows that after multiplication of \re{trans}
by $\tau_1$, \re{rohreqbar} 
will result if we define 
$\ubar$, $\fbar$, and $m$ as in equations \re{ubardef}, \re{fbar},
and \re{m} above.
\section{Appendix 3: Comment on the analysis of (Moniz and Sharp, 1977)
of a non-relativistic version of the DD equation} 

(Rohrlich, 1997) states the following, concerning the DD equation
and a non-relativistic simplification of it which will be discussed
below:
\begin{quote}
``Returning to the overview of classical charged particle dynamics,
one can summarize the present situation as very satisfactory:
for a charged sphere there now exist equations of motion
both relativistically [this refers to the DD equation \re{rohreq1}]
and nonrelativistically that make sense and that are free of
the problems that have plagued the theory for most of this century;
these equations have no unphysical solution, no runaways, and 
no preaccelerations.''
\end{quote}
This resolves into the following claims:
\begin{description}
\item[Claim 1:]
The DD equation has no preaccelerative solutions;
\item[Claim 2:]
The DD equation has no solutions which are runaway (in the future);
\item[Claim 3:] 
The DD equation has no ``unphysical solution''.
\end{description}
The analysis of this paper shows that all 
of these claims are at best optimistic and at worst false.

We showed that Claim 1 is false as stated.
However,
we also noted that it can be reformulated (by adjoining
appropriate generalized initial conditions)
so as to become
true for the special case of an field applied for a finite
time.
It is unknown whether the claim can be repaired for arbitrary fields,
as discussed in Section 3. 

We showed that Claim 2 is true for a particle moving in one space 
dimension under a force which eventually vanishes; 
it is unknown whether it is true for general
three-dimensional motion, as discussed in Appendix 1. 
Since the proof we gave for one dimension was nonroutine
and special to that dimension, 
we suspect that new ideas will be required
for a three-dimensional proof.

The truth of Claim 3 depends on what one means by ``unphysical''
solutions, but if one considers both preacceleration 
and postacceleration ``unphysical'',
then we showed that this claim cannot be repaired.

The only evidence for any of these claims offered by (Rohrlich, 1997)
or (Rohrlich, 1999) is an analysis by (Moniz and Sharp, 1977) 
of a non-relativistic version of the DD equation.
Since the non-relativistic version is an entirely different equation, 
even if the claims were true for the non-relativistic version,
they would not imply corresponding claims for the DD equation.
However, a careful reading of (Moniz and Sharp, 1977) reveals that
their analysis does not even prove the claims for their nonrelativistic
equation.

The following passage from (Moniz and Sharp, 1977)
has sometimes been interpreted 
as making Claims 1 and 2 for their nonrelativistic version
of the DD equation: 
\begin{quote}
[From the last paragraph of Section II, p.\ 2856]
``Summarizing, we have found that including the effects of radiation
reaction on a charged spherical shell results neither in runaway
behavior nor in preacceleration if the charge radius 
of the shell $L > c\tau \ldots$'' 
\end{quote}
This quote is from the part of the paper which treats only 
classical charged particles (as opposed to the quantum-mechanical treatment
of later sections).  

It is not clear whether Moniz and Sharp 
intended to assert that {\em no} solution
can be either runaway or preaccelerative (i.e., Claims 1 and 2), 
or that 
given ordinary initial conditions specifying the position and velocity
at a given time, {\em there exists} a solution satisfying these conditions
and which is neither runaway nor preaccelerative. 
The aim of the present appendix is to dispel the confusion over these
various claims by analyzing precisely what (Moniz and Sharp, 1977) 
does prove.  
We shall see that
what they actually show is closer to the latter than the former. 

The condition $L > c \tau$ corresponds in our notation to the 
condition that our delay parameter $\tau_1$ must not be
too small. 
More specifically, the condition $L > c\tau$ translates into a condition
\beq
\lbl{delaycond}
\tau_1 > \delta 
\q,
\eeq
where $\delta$ is a certain positive parameter whose exact value will not
be important to us.
Henceforth we assume this condition.
(Incidentally, for other values of the parameter $\tau_1$,
Moniz and Sharp do find unphysical solutions, either future runaway
or oscillatory.)

\newcommand{\bu}{{\bf u}}
\newcommand{\bbf}{{\bf h}}
\newcommand{\ba}{{\bf u}_0}
Their nonrelativistic version of the DD equation is:
\beq
\lbl{nonrelDD}
\frac{d\bu}{dt} = \bbf(t) - b [ \bu (t - \tau_1) - \bu(t)]
\q,
\eeq
where $\bu (t)$ represents the particle's three-dimensional velocity
at time $t$, $\bbf$ is a force-like term  (a three-dimensional force
divided by certain constants), 
and $b$ is a constant. 
This is their equation (2.10) on p. 2853 written in notation
closer to ours.  It can be obtained from the space part of the DD equation
\re{rohreq1} or \re{rohreq} by neglecting terms of quadratic or 
higher order in the velocity.

It is clearly much simpler than the DD equation. 
In particular, it is what one might call a ``linear'' delay-differential
equation, and this is critical to their analysis via Fourier transforms.

Moniz and Sharp's analysis of the possibility of solutions which 
are runaway in the future assumes that $\bbf$ eventually vanishes, 
so that for sufficiently large times, it may be dropped 
from the right side of \re{nonrelDD}, obtaining 
\beq
\lbl{nonrelDD2}
\frac{d\bu}{dt} =  b [ \bu (t - \tau_1 ) - \bu(t)]
\q.
\eeq 
This equation admits exponential solutions 
\beq
\lbl{expsol}
\bu(t) := \ba e^{\alpha t}
\q,
\eeq 
with $\ba$ a constant vector and $\alpha$ a complex constant satisfying 
$\alpha = b [e^{-\alpha \tau_1} - 1]$.  
Moniz and Sharp then show that for $\tau_1$ satisfying \re{delaycond},
the real part of $\alpha$ is negative, so that these exponential  
solutions decay to 0 as $t$ becomes large. 

That is all they prove.  But this only proves that no 
solutions of the simple {\em exponential} form \re{expsol} can be runaway
in the future, which is not the same thing as proving
that {\em no} solution can be runaway in the future. 
There are many solutions to \re{nonrelDD2} which are not exponential---indeed,
the space of all solutions is infinite dimensional (as noted
in our Section 3 and Appendix 1), whereas the space of exponential
solutions only has dimension 3.  
Thus Moniz and Sharp's analysis proves neither that no solution
of their equation \re{nonrelDD}
is runaway (Claim 2) nor that there exists a non-runaway solution
satisfying an arbitrary initial condition $\bu(t_0) = {\bf u}_0$.

(Moniz and Sharp, 1977) does give a prescription for 
writing down a formal Fourier transform of 
a solution of \re{nonrelDD} for each forcing function $\bbf$.
(They do not address the question of whether the formal expression produced, 
which contains singularities,
actually is the distributional Fourier transform of a solution.)
This prescription produces solutions 
which are not preaccelerative (in the precise sense defined in Section 3). 
Thus they show 
(to the standard of 
rigor common in physics journals like {\em Physical Review}, not to 
mathematical standards)
that {\em there exist} non-preaccelerative solutions. 

However, their proof is incomplete in one important respect---they
do not show that there exists a nonpreaccelerative solution satisfying an
arbitrary initial condition $\bu (t_0) = \bu_0$.  The nonpreaccelerative
solution $\bu$ which they produce is {\em unique}, leaving no
room to satisfy general (ordinary) initial conditions.  

Nevertheless, it seems possible that this proof could be completed. 
Even if so,
this would not prove Claim 1's assertion 
that {\em no} solutions are preaccelerative, but they may not have
intended to assert Claim 1. 

Indeed, there is an simple, explicit counterexample to Claim 1
for the nonrelativistic DD equation \re{nonrelDD}. 
Consider the case in which the force vanishes identically, i.e., 
$\bbf \equiv 0$ in \re{nonrelDD}.  For this case, {\em any}
solution for which the acceleration $d\bu /dt$ does not vanish identically
is preaccelerative---the particle accelerates {\em before} the force
is applied (because the force is never applied).  So, 
the exponential solutions $u(t) = \ba e^{\alpha t}$ previously noted in \re{expsol} 
are preaccelerative when $\ba \neq 0$, in contradiction to Claim 1. 
\section{Appendix 4: Referees' reports; Part 1:  History)}
This paper was first submitted to the {\em Journal of Mathematical Physics}.
The first referee's report was superficial and demonstrably incorrect
in some important respects, so I requested a second referee.
The second referee recommended rejection 
on the sole grounds that the problem
which both (Rohrlich, 1999) and my paper addresses
(the problem of finding a sensible equation of motion for a classical
charged particle) is  not ``relevant for physics''. 

The first referee only objected to the methods of the paper, not to 
its relevance. The second referee, who was presumably furnished
the first referee's objections to the paper's methods, 
did not question the paper's methods.

I don't agree with the second referee's value judgment, 
but I can understand and respect it. 
His report was thoughtfully written and showed understanding of the paper.
It is the only report of six (see below) which I can respect.

I had submitted the paper to {\em J. Math. Phys.} because of its mathematical
content, some of which is at a higher level than is typically
published by journals like {\em Physical Review}.
After the rejection by {\em J. Math. Phys.}, I decided to submit it
to {\em Phys. Rev. D} (PRD), which had published (Rohrlich, 1999).
I thought (naively, it turned out), that having recently 
published Rohrlich's paper on the same topic, they would find it inconsistent
to maintain that the problem which both papers address was physically
irrelevant.

The story to follow may not always reflect well on the standards of PRD, so
before giving it, I'd like to express my appreciation for the efficiency
of their online submission procedure.  
The mechanics of dealing with them is a pleasure, 
compared to most journals. 

They are efficient and fast.
Nobody likes a rejection, but an  immediate rejection is 
infinitely more courteous than
a rejection after several years of unanswered or inadequately answered
correspondence. With other journals,
the latter happens more often than it should.
(I should also remark that {\em J. Math. Phys.} was also efficient.)

Initially, PRD sent the paper to an anonymous referee
and to Professor Rohrlich as an identified referee, the latter because
the paper commented on his. 
Professor Rohrlich thought that the paper's analysis was invalid
because it replaces his original equation \re{rohreq1} with
the equivalent equation \re{rohreq}.  
I wrote him privately spelling out the equivalence (essentially 
the same analysis as Appendix 2, which was added later),
but he maintained his objection.

The anonymous referee submitted a short, superficial report
giving the impression that the paper merely quoted ``existing theorems'',
and objecting that its subject was ``interesting for the 
foundations of physics but not for phenomenology''.
(I wonder what he'd say about string theory!) 

I could understand this referee's value judgment 
if PRD were an experimental journal, but it seems inconsistent
with its publication of Rohrlich's paper in 1999 and the continued
publication of papers on the same topic since that time.
I wondered if this referee had actually read the paper 
(there was no internal evidence of that in the review),
and I wondered if he might be looking for excuses to avoid
dealing with its mathematics.

Professor Rohrlich's objection to replacement of the equation \re{rohreq1}
as he originally wrote it with the equivalent version \re{rohreq}
seemed so far out in left field that I felt I couldn't let it pass.
So, I wrote the Editor explaining why I thought that Rohrlich had made 
a serious mistake, and was continuing to make it.

He sent the paper, including Appendix 2 spelling out the
equivalence of \re{rohreq1} and \re{rohreq}, 
to a second anonymous referee (the ``Third Referee'',
including Rohrlich).   Third Referee reported
that he ``completely agree[s] with Professor Rohrlich's critique
that `results based on equation \re{rohreq} are physically 
meaningless $\ldots$'\ {''}. 

I was flabbergasted, given that the issue in dispute is so elementary,
and had been explained so carefully in Appendix 2.
The report of Third Referee gave no indication that he had even read
Appendix 2.  Surely, if there were some error in the short Appendix 2,
a conscientious referee should have pointed it out.

By this time, I was so disgusted that it was hard not to simply
walk away from the matter, but I eventually decided that for completeness
and closure, I should protest this latest incompetence. 
So, I wrote the Editor, D. Nordstrom, explaining that I believed 
that, however unlikely it might seem, 
Third Referee's report was also seriously in error.

I suggested the following way of resolving the matter to the satisfaction
of all:  submit the narrow issue of the equivalence of \re{rohreq1}
and \re{rohreq} to an identified referee acceptable both to
PRD and me.  (I suggested Barry Simon, whom I have never met,
but whose competence I respect, based on his published work.)   
I promised that if such a referee ruled that these two equations were
not equivalent, I would immediately withdraw the paper and 
never submit another to PRD. 

He didn't reply to that, but he did send the paper back 
to Third Referee, who reversed himself, writing:
\begin{quote}
``I have no doubt about the mathematical equivalence of equs. 
(6) and (7).  (And I have read and confirmed the calculations
in Appendix 2!)''
\end{quote}
But this referee maintained his recommendation of rejection
on the basis that (a) ``the topic of a classical equation
of motion does not belong to the most urgent questions
in theoretical physics $\ldots$'', and (b)  
the paper does not address the issue of the size of the postacceleration
predicted by the DD equation.
\section{Appendix 4, Part 2: All of the referees'  reports, in full}
This section of
the previous version, titled as above, 
contained all of the referees' reports,
along with discussions of them. 
The verbatim reports have been deleted from this version 
at the request of the arXiv.  

I am not convinced that it was legally necessary to remove the 
referees' reports.  
The legal issues involved seem to me obscure and  unsettled.  
I shall not attempt to summarize them here.

From a commonsense standpoint, it's hard to imagine why inclusion
of the verbatim reports
should be less acceptable than paraphrasing them 
(which is undisputedly allowed).  
If it should turn out that this is indeed what the law requires,
then the law should be changed. 
The only reason for including the reports verbatim was to 
avoid any possible distortion of the referees' meanings.

Some of the six reports were already paraphrased in the preceding section.
To paraphrase them all and rewrite the discussions accordingly would
require more work than I am willing to do.

I will furnish the previous version, including the verbatim reports,
to anyone interested, by email or other means.  
For further information, visit www.math.umb.edu/$\sim$sp.
\section{Appendix 5:  Solving the DD equation backwards in time}
The referees for {\em Physical Review D} (PRD) raised the following
objections (and only these) to the present paper.
\begin{enumerate}
\item
The version of the DD equation analyzed by the paper, \re{rohreq},
is not physically equivalent to the one proposed by Rohrlich, \re{rohreq1}.
\s
After many months and many pages of careful explanation,
the only independent referee to affirm this objection
finally withdrew it.  So, one hopes that this issue may be regarded as 
settled. 
\item
The subject of the correct equation of motion for classical charged
particles is of insufficient physical interest.
\item
The paper is too mathematical.
\item
The paper does not estimate the size of the postaccelerations 
predicted by the DD equation.
\end{enumerate}
Objection 2 seems odd, given that PRD recently published Rohrlich's
paper \cite{rohrlichpr} on precisely this topic 
(equations of motion for classical charged
particles), along with papers on the same topic by other authors.
However, taking it at face value, it occurred to me that it certainly shouldn't
apply to a ``Comment'' paper, e.g., ``Comment on {\em Classical self-force} 
by F. Rohrlich''.  (If it did apply, that would amount to an editorial
declaration that the subject of {\em Classical self-force} was of insufficient
physical interest to merit publication, not long after PRD had published it!) 
Also, such a paper, (which is required to be short
and therefore could not contain a full mathematical analysis) 
would obviate Objection 3.

I wrote to Editor D. Nordstrom asking if PRD would consider such a ``Comment''
paper, or if perhaps the editors had decided not to publish any form
of the present work.
After he didn't reply, I rewrote the paper 
as a Comment paper and submitted it on July 28, 2004.
Its submission was acknowledged immediately, but by the end of December,
I had heard nothing more.  I wrote on December 28
enquiring about the status of the paper, and in particular if
it had been sent to any referee.  On January 28, 2005, I received
a cryptic reply apologizing that the paper had been ``misfiled''.
and stating that it had been ``sent out for review''. 
(This is being written on January 31.)

The presentation of the Comment paper is based on the fact
that the solution manifold for the DD equation is {\em infinite-dimensional},
rather than of finite dimension as expected (cf. Appendix 1).
Thus to base a physically reasonable theory on this equation,
one needs some auxiliary condition to reduce the solution manifold
to finite dimension.

That the DD equation's solution manifold is of infinite dimension
is implicit in the present work and strongly suggested by the fact that
the space of generalized initial conditions is obviously infinite-dimensional.
However, no careful proof that the solution manifold is infinite-dimensional
was included above because it was only used for motivation
in peripheral discussions.

The ``Comment'' paper contains a broad mathematical outline,
but no substantial proofs, 
for which  the reader is referred to the present work. 
After submitting it, I realized that the present work didn't give  
a real proof that the solution manifold of the DD equation 
is infinite-dimensional.  
This appendix remedies this by furnishing a proof.  

Although the Comment paper uses Rohrlich's original form \re{rohreq1}
of the DD equation,
for notational simplicity, 
we use here the normalized DD equation \re{rohreq}, so that 
the time units are normalized to make the delay $\tau_1$ equal
to 1:
%\beq
%test of equation counter
%\eeq
\newcounter{tempsave}
\setcounter{tempsave}{\value{equation}} 
\setcounter{equation}{\value{rohreqnum}}
\addtocounter{equation}{-1}
\beq 
 m \frac{du^i}{d\tau} = f^i (\tau) + 
 [ u^i(\tau-1) - u^\alpha (\tau - 1) u_\alpha (\tau) u^i(\tau) ] 
\q,
\eeq 
\setcounter{equation}{\value{tempsave}}
%\beq
%test of equation counter
%\eeq
Recall that a generalized initial condition is a specification of $u$
on some interval $\tau_0 - 1 \leq \tau \leq \tau_0$ of length 1.
(For a general delay $\tau_1$, the interval of specification would
be an interval of length $\tau_1$.)
Section 3 noted that to every such generalized initial condition corresponds
a unique solution defined on $[\tau_0, \infty)$.
In other words, the DD equation can be solved {\em forward} in time starting
with any generalized initial condition.

We are now going to show how to solve the DD equation {\em backward} in time 
to obtain a unique solution on $(-\infty, \tau_0)$ for each generalized
initial condition.   
In Section 3, this was done in a simple, {\em ad hoc} way 
for the DD1 equation, which was all that was needed for the treatment given.
But the Comment paper doesn't mention the DD1 equation,
so now we need to do it for the full DD equation \re{rohreq}.

Combining the forward solution with the backward solution
gives a unique solution on $(-\infty, \infty)$ 
corresponding to each generalized initial condition on $(\tau_0 - 1, \tau_0)$. 
We define the solution manifold to be the set of all solutions
defined on $(-\infty, \infty)$.

Then it becomes obvious that the solution manifold is
parametrized by the set of all generalized initial conditions
on any given interval $(\tau_0 - 1, \tau_0)$ of length 1.
The technical issue is whether the assignment of a solution to 
each generalized initial condition is one-to-one (hence bijective).
This is obvious if ``solutions'' are required to be defined on 
$(-\infty, \infty)$ because then each generalized initial condition
is the restriction of a solution. 
(It's not obvious if ``solutions'' are only required
to be defined on $(\tau_0, \infty)$, though it can be proved.) 
 
To see how to solve the DD equation backward in time starting
from a generalized initial condition, 
attempt to solve in \re{rohreq} for $u(\tau - 1)$ in terms of $u(\tau)$,
$du/d\tau $, and $f(\tau)$.  This is impossible because
the bracketed quantity is only the projection of 
$u(\tau - 1)$  on $u(\tau)$, 
but the equation does uniquely determine this projection.
Hence it uniquely determines $u(\tau - 1)$ up to an additive term 
$\beta u(\tau)$, $\beta$ a scalar:
$$
u(\tau - 1) = \beta u(\tau) + 
 [m \frac{du^i}{d\tau} - f^i (\tau) ] = \beta u(\tau) + q(\tau)
\q,
$$
where for brevity we introduce the name 
$$q(\tau) := [m \frac{du^i}{d\tau} - f^i (\tau) ]
$$ for the bracketed quantity. 

\newcommand{\bq}{{\bf q}}
Routine calculation reveals that
the scalar $\beta $  is uniquely determined by the necessary condition
that the four-velocity $u(\tau - 1)$ must be a future-pointing unit vector. 
The condition that $u(\tau-1)$ have unit norm gives a quadratic equation
for $\beta$, and the condition that $u(\tau - 1)$ be future-pointing
singles out a unique solution to this quadratic equation.
The result of this calculation is: 
$$
u(\tau - 1) = [-u^\alpha (\tau) q_\alpha (\tau)+ 
 \sqrt{ 1 + (u^\alpha(\tau) q_\alpha(\tau))^2 - 
(q^\alpha(\tau) q_\alpha(\tau))^2 } ] u(\tau) +  
q(\tau)
.
$$

Thus given the four-force $f$, 
the values of $u(\tau)$ on an interval $[\tau_0, \tau_0 + 1]$,
uniquely determine the values of $u(\tau - 1)$ on this interval,
Equivalently stated, the values of $u(\tau)$ on $[\tau_0, \tau_0 + 1]$
uniquely determine the values of $u(\tau)$ on $[\tau_0 - 1, \tau_0]$,
and, by iteration, the values of $u(\tau)$ on $(\infty, \tau_0]$ 
 
%MARKEND

%

\begin{thebibliography}{00}
%
\bibitem{caldirola} 
Caldirola, P., ``A new model of classical electron'',
Nuovo Cimento {\em 3}, Suppl. 2, 297-343 (1956) 
%
\bibitem{chicone} C. Chicone,  
Inertial and slow manifolds for delay equations with small delays,
\emph{J. Diff. Eqs.}, {\bf 190} (2003), 364--406.
%
\bibitem{driver} Driver, R.D.,  
On Ryabov's asymptotic characterization of the solutions of
quasi-linear differential equations with small delays, \emph{SIAM Review}
{\bf 10}(3)  (1968), 329--341.  
\bibitem{eliezer}
Eliezer, C.J., ``The hydrogen atom and the classical theory of radiation'',
Proc. Camb. Phil. Soc. {\bf 39}, 173ff (1943)  
%
\bibitem{eliezer2}
Eliezer, C.J., ``On the classical theory of particles'', 
Proc. Royal Soc. London {\bf A194}, 543-555 (1948)  
\bibitem{ms}
Moniz, E.\ J., and Sharp, D.\ H.,  
``Radiation reaction in nonrelativistic quantum electrodynamics'',
Phys.\  Rev.\ {\bf 15}, 2850-2865 (1977)
\bibitem{parrottbook}
Parrott, S., {\em Relativistic Electrodynamics and Differential Geometry},
Springer, New York, 1987
\bibitem{parrott1} Parrott, S.,
Radiation from a uniformly accelerated charge and the equivalence principle,
Found.Phys. {\bf 32} (3), 407-440 (2002)
\bibitem{page}
Page, L., 
``Is a moving mass retarded by the reaction of its own radiation?'',
Phys. Rev. {\bf 11}, 377-400, 1918 
\bibitem{perko}
Perko, L., {\em Differential Equations and Dynamical Systems},
Second Edition, (Springer, N.Y., 1996) 
\bibitem{rohrlichbook}
Rohrlich, F., {\em Classical Charged Particles}, Addison-Wesley,
Reading, MA, 1965 
\bibitem{rohrlichamj}
Rohrlich, F., ``The dynamics of a charged sphere and the electron'',
Am. J. Phys. {\bf 65} (11), 1051-1056 (1997)
\bibitem{rohrlichpr}
Rohrlich, F., ``Classical self-force'', 
Phys. Rev. D {\bf 60}, 084017 (1999)  
%
\bibitem{ryb1} Ryabov, Yu. A,  
Application of the method of small parameters to the investigations
of automatic control systems with delay,
\emph{Automat. Remote Control} {\bf 21} (1960), 507--514.
\bibitem{ryb2} Ryabov, Yu A., 
Application of the method of small parameters for the construction of 
solutions of differential equations with retarded argument,
\emph{Sov. Math. Dokl.} {\bf 1} (1960), 852--855.
\bibitem{ryb3}  Ryabov, Yu. A., 
Application of the method of small parameters of Lyapunov-Poincar\'e
in the theory of systems with delay, 
\emph{Inzh. Zh.} {\bf 1} (1961), 3--15.
\bibitem{ryb4} Ryabov, Yu A., 
Certain asymptotic properties of linear systems with small time delay,
\emph{Sov. Math. Dokl.} {\bf 4} (1963), 928--930.
\bibitem{ryb5} Ryabov, Yu A., 
Certain asymptotic properties of linear systems with small time lag,
\emph{Trudy Sem. Teor. Differential. Uravnenii s Otklon. Argumentom Univ.
%
Druzhby Narodov Patrisa Lumumby} {\bf 1} (1965), 153--164.  
\bibitem{Yaghjian}
Yaghjian, A.\ D., {\em Relativistic Dynamics of a Charged Sphere},
Springer-Verlag, Berlin, 1992  
\end{thebibliography}
\end{document}